\documentclass[apjl]{emulateapj}

\def\ltsima{$\; \buildrel < \over \sim\;$}
\def\ltsim{\lower.5ex\hbox{\ltsima}}
\def\gtsima{$\; \buildrel > \over\sim \;$}
\def\gtsim{\lower.5ex\hbox{\gtsima}}
\def\ms{$M_{\odot}$ }
\def\msp{$M_{\odot}$}

\slugcomment{Accepted for publication in ApJ Letters}

\begin{document}
\title{Relics of metal-free low mass stars exploding as thermonuclear supernovae}

\author{Takuji Tsujimoto$^1$ and Toshikazu Shigeyama$^2$}

\altaffiltext{1}{National Astronomical Observatory, Mitaka-shi,
Tokyo 181-8588, Japan; taku.tsujimoto@nao.ac.jp}
\altaffiltext{2}{Research Center for the Early Universe, Graduate School of Science, University of Tokyo, 7-3-1 Hongo, Bunkyo-ku, Tokyo 113-0033, Japan; shigeyama@resceu.s.u-tokyo.ac.jp.}

\begin{abstract}
Renewed interest in the first stars that were formed in the universe has led to the discovery of extremely iron-poor stars. Since several competing scenarios exist, our understanding of the mass range that determines the observed elemental abundances remains unclear. In this study, we consider three well-studied metal-poor stars in terms of the theoretical supernovae (SNe) model. Our results suggest that the observed abundance patterns in the metal-poor star BD $+80^\circ$245 and the pair of stars HD 134439/40 agree strongly with the theoretical possibility that these stars inherited their heavy element abundance patterns from SNe initiated by thermonuclear runaways  in the degenerate carbon-oxygen cores of primordial asymptotic giant branch stars with masses of $\sim 3.5-5$ \msp. Recent theoretical calculations have predicted that such SNe could be originated from metal-free stars in the intermediate mass range. On the other hand, intermediate mass stars containing some metals would end their lives as white dwarfs after expelling their envelopes in the wind due to intense momentum transport from outgoing photons to heavy elements. This new pathway for the formation of SNe requires that stars are formed from the primordial gas.  Thus, we suggest that stars of a few solar masses were formed from the primordial gas and that some of them caused thermonuclear explosions when the mass of their degenerate carbon-oxygen cores increased to the Chandrasekhar limit without experiencing efficient mass loss.
\end{abstract}


\section{Introduction}

Elemental abundance of recently discovered extremely iron-deficient stars has raised questions about the mass range of the first stars that were formed in the universe \citep{Christlieb_02, Umeda_03, Schneider_03, Shigeyama_03, Suda_04, Frebel_05, Iwamoto_05}. General theoretical considerations seem to suggest that, due to inefficeint cooling of the primordial gas, the first stars were at least as massive as 100 \ms  \citep[e.g.,][]{Abel_02,Bromm_02}. However, other studies claim that molecular or atomic hydrogen cooling is sufficient to form low-mass stars from the primordial gas \citep{Nakamura_01,Omukai_03}. Therefore, the detection of signatures constraining the mass range of the first stars limits the possible formation mechanisms of these objects.

The key parameters controlling the structure and evolution of stars are mass and metallicity. A given combination of these two parameters establishes new pathways for supernova (SN) formation from  {\it ``primordial intermediate mass"} stars. According to recent calculations \citep{Fujimoto_00, Suda_04}, stars with [Fe/H] \ltsim $-5$ and masses in the range of 3.5\ms \ltsim $M$ \ltsim 7 \ms do not experience mixing episodes including the third dredge-up, in their convective envelope in the thermally pulsing asymptotic giant branch (TPAGB) phase. These stars will end up with a small amount of carbon as well as $s$-process elements in their envelopes. As a result, these stars are less opaque and more compact than stars that have experienced the third dredge-up. These features also result in stars with lower radiation pressures and stronger surface gravities. Therefore, it is expected that the mass loss from an extremely metal-poor star will be substantially suppressed \citep{Fujimoto_84}. Thus, a degenerate carbon-oxygen core can reach the Chandrasekhar limit  ($M_{\rm ch}$) while the star is in the TPAGB phase.  Eventually the star undergoes a thermonuclear explosion to become an SN. This type of SN can be classified as a type Ia in terms of nucleosynthesis but as a type II in terms of spectral type. If the mass of a proto-galactic cloud is $10^6$ \ms corresponding to the Jeans mass soon after recombination, a typical Type II SN (SN II)  enriches the cloud up to [Fe/H]$\sim -4$. Therefore only the primordial (population III) stars are likely to end up as such SNe. Hence, we define them as SNe IIIa.

\begin{figure*}[t]
\begin{center}
\includegraphics[width=13.5cm,clip=true]{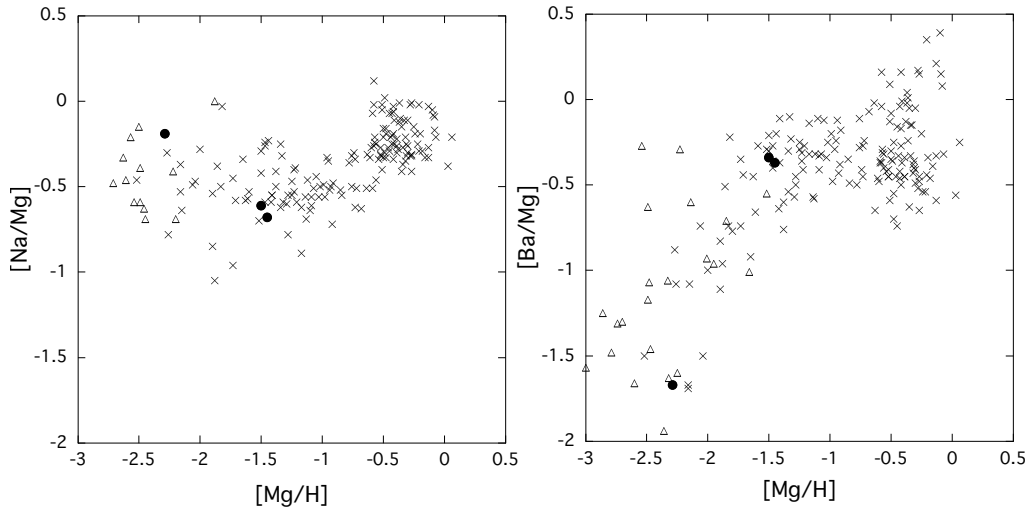}
\end{center}
\caption{{\it Left panel}: Correlation of [Na/Mg] with [Mg/H] for BD+80$^\circ$245 and HD 134439/40 \citep[filled circles;][Fulbright 2000]{Ivans_03} together with other Galactic stars \citep[open triangles;][crosses; Fulbright 2000]{Cayrel_04}. No abundance anomaly is seen for the three stars. {\it Right panel}: The same as left panel but for [Ba/Mg]. The data shown by open triangles are taken from \citet{McWilliam_98}.}
\end{figure*}

The lifetime of an SN IIIa progenitor is much shorter than that of an SN Ia progenitor, though these two types of SNe supply similar amounts of nuclear products. In a white dwarf, an SN Ia event needs substantial replenishment of the gas by accretion from a low-mass companion. As a result, SN Ia progenitors have much longer evolutionary time-scales than those of SNe II $-$, typically longer than $10^9$ yrs $-$. This prevented SNe Ia from contributing to chemical enrichment in the early epoch of galaxy evolution. On the other hand, an SN IIIa event occurs when an intermediate mass star evolves into the AGB phase in a period ranging from several $10^7$ yrs to a few $10^8$ yrs, depending on the progenitor mass. Such short time scales could enable us to detect evidence of the above in the elemental abundances of old stars. Recent studies of the chemical composition of metal-poor stars have revealed that they have inherited the abundance pattern of the ejecta of the preceding SNe \citep[e.g.,][]{Audouze_95, Shigeyama_98, Umeda_03}. If this is the case, there might be some later stars that inherited the nuclear products of SNe IIIa, though their abundance patterns could be modified to some degree by products from SNe II.

The most characteristic abundance feature of stars inheriting the nuclear products of an SNe IIIa event is expected to be the deficiency of $\alpha$-elements, compared with the enhanced $\alpha$/Fe ratios commonly seen in halo stars. In fact, there exist so-called low-$\alpha$ stars in the Galactic halo with low $\alpha$/Fe ratios. It has been proposed \citep{Nissen_97, Gilmore_98} that these low-$\alpha$ stars were born from accreted dwarf galaxies or protogalactic fragments, where a prolonged low star formation rate enables SNe Ia events to occur in a low-metallicity regime. The specific mechanism which explains the low $\alpha/$Fe ratios in an individual star is easily determined by the abundance of $s$-process elements, since the time scales associated with these two types of SNe are substantially different. The long lifetime of SNe Ia allows AGB stars to substantially pollute the gas with $s$-process elements before low-$\alpha$ stars are formed, whereas SNe IIIa preceded most AGB stars with masses \ltsim$3.5\,M_\odot$. 

In the literature, we found three stars (BD+80$^\circ$245 and HD 134439/40) that exhibit the abundance patterns expected from SNe IIIa. Elemental features of these three stars are discussed in the subsequent two sections. Their origin is investigated in the context of a chemical evolution model developed by \citet{Tsujimoto_99} in \S \ref{ee}.

\section{Elemental features of BD+80$^\circ$245}

The conspicuous abundance feature of the star BD $+80^\circ$245 is the deficiency of  $\alpha$-elements and $n$-capture elements, such as Ba or Eu, and low ratios of Na, Al, Sc, Zn relative to Fe \citep{Fulbright_00, Stephens_02,Ivans_03}. In the supernova-induced star formation scenario proposed by \citet{Shigeyama_98}, the stellar abundance pattern is determined by the combination of heavy elements ejected from an SN itself and those elements that are already present in the interstellar gas due to previous SN II explosions \citep{Tsujimoto_99}. Therefore, the deficiency not only of $\alpha$-elements but also of $n$-capture and odd-numbered elements should be attributed to an additional Fe supply from an SN IIIa. Since an SN IIIa produces few $\alpha$-elements, if we take the abundance ratios of these elements to Mg instead of Fe, both BD+80$^\circ$245 and HD 134439/40 become indistinguishable from other halo stars (see Fig.~1). In other words, the abundance of the interstellar medium (ISM) [Mg/H]$_{\rm gas}$ at the birth epoch of this star must be equal to the observed abundance of BD+80$^\circ$245, [Mg/H]$=-2.29$  \citep{Ivans_03}. Assuming that the abundance ratio is equal to the average ratio for halo stars [Mg/Fe]$_{\rm gas}$=0.4, the Fe abundance in the ISM is estimated to be [Fe/H]$_{\rm gas}$=$-2.69$. Adding another $0.63$\ms of the Fe supplied by the preceding SN IIIa to the $6.5\times10^4$ \ms of the gas swept up by the SN remnant \citep{Shigeyama_98}, the metallicity of a star formed from this gas becomes [Fe/H]=$-1.97$. This is in good agreement with the observed value of [Fe/H]=$-2.07$ \citep{Ivans_03}.

Another reason for connecting stellar abundance features with a Pop III SN IIIa is the deficiency of odd-numbered elements (Mn, Co) compared to even-numbered ones (Cr, Ni) among the Fe-group. This feature may favor nucleosynthesis in primordial SNe, since odd nuclear charge numbered elements require the pre-existence of heavy element seeds in the SN progenitor. Among these Fe-group elements, Mn abundance best distinguishes the products of SNe IIIa from those of SNe Ia. If an SN Ia, instead of an SN IIIa, supplied the Mn and the Fe to the gas from which this star formed, then a theoretical SN Ia model \citep{Iwamoto_99} would predict that the [Mn/Fe] ratio in the gas, which was initially $\sim -0.4$, would become $\sim-0.07$ after being swept up by the SN. In fact, the observed [Mn/Fe] =$-0.26$ \citep{Ivans_03} implies that the Mn yield from an SN IIIa is about one-third that of an SN Ia. Likewise, the ejecta of an SN Ia would result in [Co/Fe]$\sim 0$, whereas observations show [Co/Fe]=$-0.18$ \citep{Ivans_03}.

\begin{figure}[t]
\begin{center}
\includegraphics[width=8cm,clip=true]{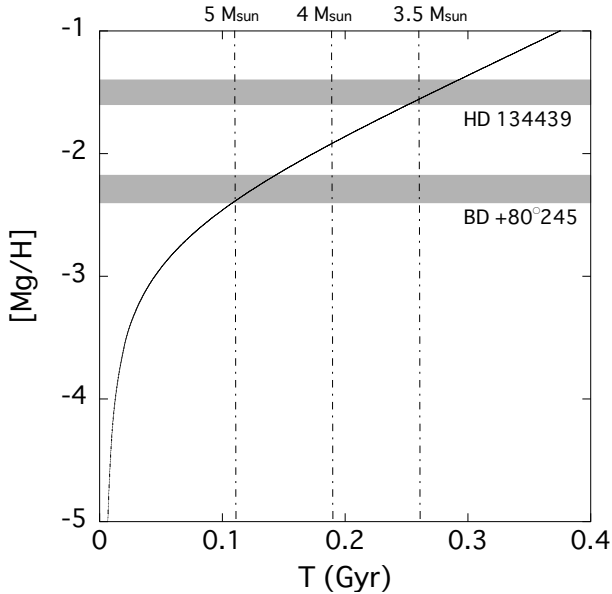}
\end{center}
\caption{The predicted age-[Mg/H] relationship for the gas in the protogalactic cloud. SN IIIa progenitor masses corresponding to different stellar lifetimes are also indicated. The shaded areas denote [Mg/H] abundances with errors for BD+80$^\circ$245 and HD 134439. The [Mg/H] abundance of HD 134440 is not shown since it only differs by 0.05 dex from that of HD 134439.}
\end{figure}

\section{Elemental features of HD 134439/40}
The pair of stars HD 134439/40 \citep{King_97} also appear to inherit the abundance pattern of the ejecta of an SN IIIa. They both exhibit the deficiency of $\alpha$- and $n$-capture elements similar to BD $+80^\circ$245, but to a lesser extent \citep{Ryan_91, Fulbright_00, Gratton_03}. This mild deficiency is naturally interpreted in the framework of the evolution of stellar abundances in the Galactic halo. The relatively small contribution of SNe IIIa to stellar abundance is attributed to a less massive progenitor star. Its longer lifetime allows more numerous SNe II to enrich the interstellar matter until the star is formed. In \S \ref{ee}, it will be shown that the unusual elemental abundances of these three stars are linked to the unique evolutionary track of stellar abundances in the ejecta of SNe IIIa with different progenitor masses.

\section{The evolution of elemental abundances of stars triggered by SN IIIa explosions}\label{ee}

\begin{figure*}[t]
\begin{center}
\includegraphics[width=14.2cm,clip=true]{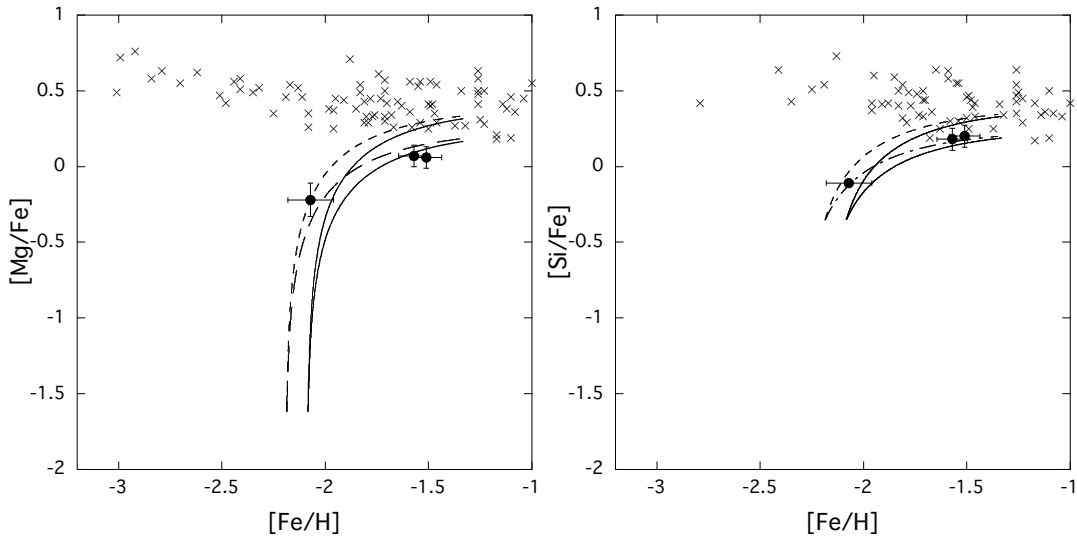}
\end{center}
\caption{{\it Left panel}: The predicted correlation of [Mg/Fe] with [Fe/H] inside the shell swept up by the ejecta of an SN IIIa for different progenitor masses, together with the observational data for BD+80$^\circ$245 and HD 134439/40 \citep[filled circles;][Fulbright 2000]{Ivans_03} as well as other Galactic stars \citep[crosses;][]{Fulbright_00}. Individual model curves show four cases with combinations of the swept-up mass and the [Mg/Fe] of the gas, that is ($8\times10^4$ \msp, 0.4)(upper dashed curve), ($6.5\times10^4$ \msp, 0.4)(upper solid curve), ($8\times10^4$ \msp, 0.25)(lower dashed curve), and ($6.5\times10^4$ \msp, 0.25)(lower solid curve), respectively. See text for details. {\it Right panel}: The same as left panel but for [Si/Fe]. No observational error for [Si/Fe] of BD+80$^\circ$245 is provided.}
\end{figure*}

In this section, we will discuss how the unusual elemental abundances seen in the three stars under discussion can be modeled theoretically. The chemical evolution considered here is based on a scenario \citep{Tsujimoto_99} in which the stellar abundances change through successive sequences of the SN explosion, shell formation, and star formation from the matter swept up by the individual SN remnants (SNRs). Heavy elements ejected from an SN are assumed to be trapped and well mixed within the SNR shell. A fraction of this heavy element material is then incorporated into the next generation stars, while the material remaining in the gas will mix with the ambient medium. The mass fraction of the first-generation stars is introduced as a free parameter in this model, and this value is assumed to be $2.5\times 10^{-4}$ \citep{Tsujimoto_00}. Details of this process are given in \citet{Tsujimoto_99}.

First, we show the evolution of [Mg/H] in the gas (Fig.~2). Since an SN IIIa supplies only a small amount of Mg, the Mg abundances of the stars introduced in the preceding sections coincide with the abundance of the interstellar gas at the birth epochs of the stars. Thus, this age-[Mg/H] relation can be used to date these stars. That is, for a given [Mg/H] value, the progenitor mass of the SN IIIa for each star can be deduced. The resultant masses are 5 \ms for BD $+80^\circ$245 and 3.5 \ms for HD 134439. These inferred masses are within the allowed mass range ($\sim 3.5-7$\msp) for SN IIIa progenitors. The lower bound mass is required for the inhibition of mixing processes in the convective envelope \citep{Fujimoto_00}, while the upper bound mass is the lower limit at which hydrostatic carbon burning will occur in the stellar core \citep{Siess_02}.

In contrast to Mg abundance, stellar Fe abundance inherited from SN IIIa remnants differs from that in the interstellar gas, which leads to unusual elemental [X/Fe] ratios for these stars. Since the amount of heavy elements contained in the gas increases with time, the major contributor to stellar metallicity switches from SN ejecta to the metallicity in the interstellar gas. Therefore, the delay in SN IIIa explosions reduces the anomalies of the [X/Fe] ratios. As a result, the [X/Fe] ratios of stars trace the continuous evolutionary track in Figure 3, which considers the assemblage of elements in the ejecta of individual SNe IIIa, with different progenitor masses, in addition to the elements in the ISM. As well, this figure shows the model curves for the evolutionary change in [Mg/Fe] (left panel) and [Si/Fe] (right panel) ratios together with the observed data for the three stars \citep{Ivans_03, Fulbright_00} and numerous other Galactic stars with {\it Hipparcos} parallaxes \citep{Fulbright_00}. For the nucleosynthetic yields of Mg, Si, and Fe in SNe IIIa, those predicted by the W7 model of SNe Ia \citep{Iwamoto_99} are adopted. A small Mg yield in SN IIIa leads to the formation of stars with an abundance ratio, [Mg/Fe], as low as 
$\sim-1.6$. In contrast, a relatively large Si yield in SN IIIa results in stars with an [Si/Fe] ratio $>-0.4$. Theoretically, the mass swept up by an SNR is dependent on the local gas density \citep{Shigeyama_98} and is found to lie within a factor of $\sim$3 about the mean of $6.5\times10^4$\ms \citep{Nakasato_00}. In this study, we consider a factor of 1.25 larger than this (dashed curves) together with the typical case (solid curves) (see Figure 3). Taking into account the incomplete mixing of elements from different SNe inside the inhomogeneous ISM, we calculated the abundance ratio of stars formed from SNe IIIa with two different values of the gaseous [Mg (Si)/Fe], i.e., 0.4 (upper two curves) and 0.25 (lower two curves). These results reveal that the observed lower [Mg/Fe] and [Si/Fe] ratios for the three stars are well described by the unified scheme. This scheme proposes that these stars were formed from the gas swept up by SNe IIIa with different progenitor masses during different epochs. 
 
Moreover, the predicted relationships between [Ba/Fe] and [Fe/H], as well as [Eu/Fe] and [Fe/H], for the three stars also agree with the observed ratios, as shown in the two panels of Figure 4. The evolution of [Ba/Fe] in the gas is calculated assuming that the production site for early-epoch $s$-process elements is $3-5$ \ms AGB stars, which is shown by the dash-dotted curve. It was assumed that the [Eu/Fe] ratio in the gas is 0.5. The predicted [Ba/Fe] and [Eu/Fe] sequences of stars formed from SNe IIIa with different progenitor masses are completely different from the evolution of gaseous abundances, as is shown by the solid and dashed curves. Thus, the remarkably low [Ba/Fe] and [Eu/Fe] ratios for BD $+80^\circ$245 are attributable to an early SN IIIa explosion in very low-metallicity gas due to a relatively massive ($\sim 5$\msp) progenitor. On the other hand, the SN IIIa progenitor, having imprinted the relatively high [Ba/Fe] and [Eu/Fe] ratios in HD 134439/40, was less massive ($\sim 3.5$\msp) than the SN IIIa progenitor forming BD+80$^\circ$245. At the same time, the model curves in Figures 3 and 4 predict the location of stars formed from SNe IIIa in the [$\alpha$/Fe]-[Fe/H] and [$n$-capture/Fe]-[Fe/H] planes. Finally it should be noted that heavy elements ejected from SNe IIIa are expected to have little influence on the overall enrichment of interstellar gas, since SNe IIIa, whose progenitors are first-generation stars, are underpopulated, and SNe II had already enriched the interstellar gas with heavy elements before the emergence of SN IIIa.

\section{Conclusions and Discussion}

\begin{figure*}[t]
\begin{center}
\includegraphics[width=14.2cm,clip=true]{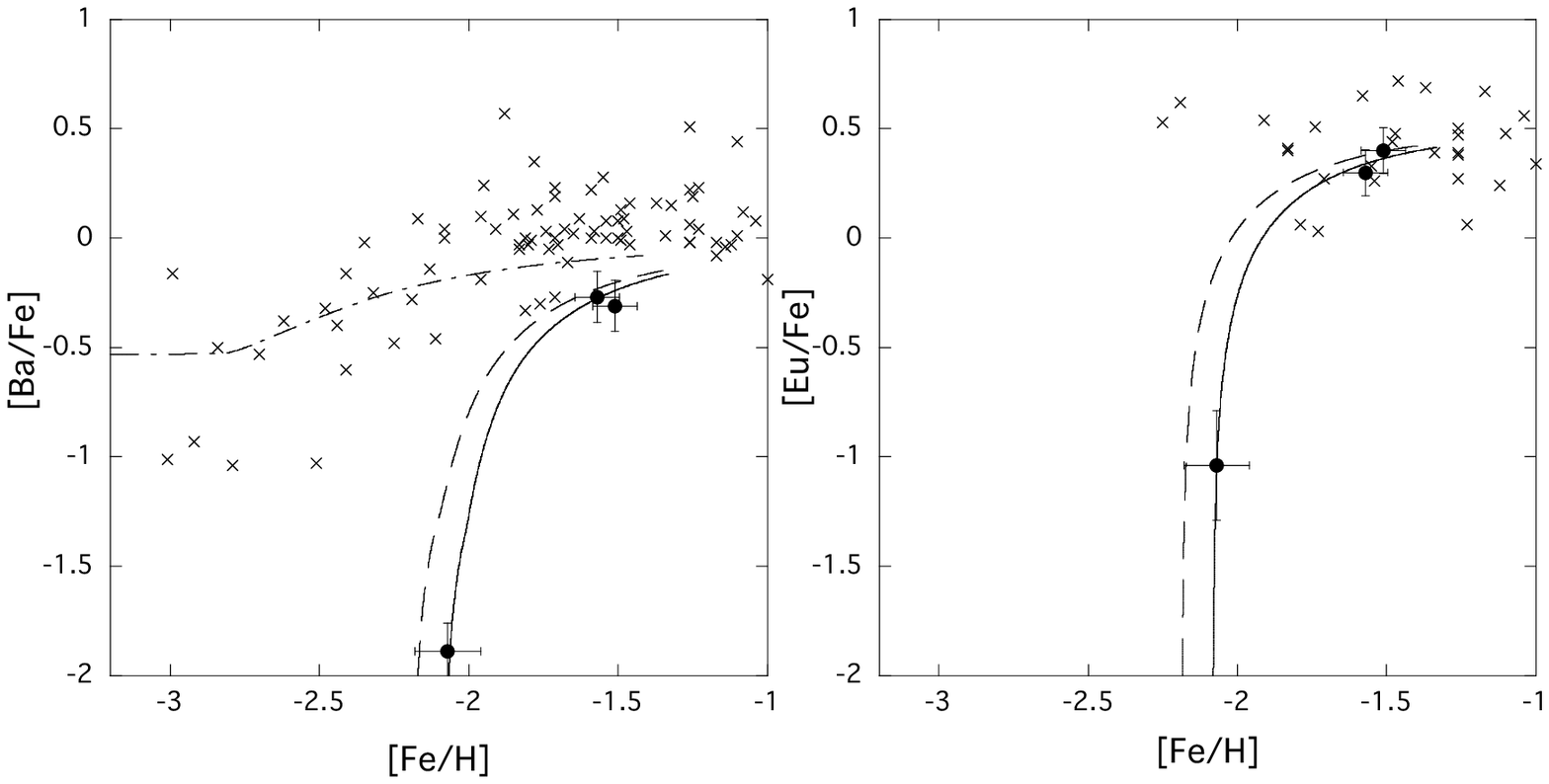}
\end{center}
\caption{{\it Left panel}: The same as Figure 3 but for [Ba/Fe]; solid and dashed curves correspond to the swept-up mass of $6.5\times10^4$\ms and $8\times10^4$\msp, respectively. Dashed-dotted line shows the evolution of the abundance ratios in the gas.  {\it Right panel}: The same as left panel but for [Eu/Fe]. The observational data for BD+80$^\circ$245 are taken from \citet{Fulbright_00} instead of \citet{Ivans_03} since \citet{Fulbright_00} gives a consistent $r$-process Ba/Eu ratio.}
\end{figure*}

The unusually low abundances of $\alpha$- and neutron-capture elements relative to Fe seen in the metal-poor star BD $+80^\circ$245 and the pair of stars HD 134439/40 are shown to be fossil records of SN explosions triggered by a thermonuclear runaway in a metal free intermediate mass star. These SNe have two distinct characteristics. First, they can be classified as type Ia in terms of nucleosynthesis, that is, they synthesize a large amount of Fe as a result of the thermonuclear explosions. Second, for their cores to grow to the Chandrasekhar limit during the TPAGB phase, the progenitors need to be metal-free stars. Since the three stars exhibit these two characteristics, ``a sufficient Fe supply on a short time-scale" and ``primordiality" in their elemental abundance patterns, this strongly suggests that there must be metal-free stars of a few solar masses.

We also predict that stars inheriting heavy elements from SNe IIIa populate distinct branches in the [$\alpha$/Fe]$-$[Fe/H] and [$n$-capture/Fe]$-$[Fe/H] planes. The manner in which these stars populate this branch should give valuable information on the initial mass function of metal-free stars. Stars descended from SNe IIIa are predicted to have metallicities in the range of $-2.3$\ltsim[Fe/H]\ltsim$-1.3$. Thus, the eight low-$\alpha$ halo stars with $-1.3\leq[{\rm Fe/H}] \leq -0.8$ \citep{Nissen_97}, and D119 with [Fe/H] =$-2.95$ exhibiting extreme deficiencies of Ca, Sr, and Ba in the Draco dwarf spheroidal galaxy \citep {Fulbright_04}, are unlikely to be associated with the proposed SNe IIIa.

Theoretically during the formation of galaxies, stars form in the protogalactic clouds assembling to construct the Galactic halo. A spread of several billion years in the stellar age among the halo stars \citep{Schuster_89} suggests that heavy elements dispersed into the halo from the protoclouds should have polluted other clouds before star formation had started there. Thus, the formation of metal-free stars must have started at a very early epoch and, accordingly, very far from the Galactic center. Thus, stars inheriting heavy elements from Pop III SNe are expected to share kinematic features of the outermost halo population, such as large apo-galactic orbital radii with high eccentricities. Observations reveal that the three stars examined in the current letter do indeed possess such kinematic features \citep{Carney_97, King_97}.

\vspace{2cm}
\acknowledgements

We are grateful to the anonymous referee for the useful comments that helped improve this paper. TS thanks Masayuki Fujimoto for the valuable conversation on the evolution of metal-free stars, which stimulated this work. We also appreciate the careful reading of the manuscript and suggestions for improvement by J. B. Carr. This work has been partially supported by a grant-in-aid (16540213) of the Japanese Ministry of Education, Science, Culture, and Sports.


\begin{thebibliography}{}
\bibitem[Abel et al.(2002)]{Abel_02}
Abel, T., Bryan, G. L., \& Norman, M. L. 2002, Science, 295, 93
\bibitem[Audouze \& Silk(1995)]{Audouze_95}
Audouze, J., \& Silk, J. 1995, ApJ, 451, L49
\bibitem[Bromm et al.(2002)]{Bromm_02}
Bromm, V., Coppi, P. S., \& Larson, R. B. 2002, ApJ, 564, 23
\bibitem[Carney et al.(1997)]{Carney_97}
Carney, B. W., Wright, J. S., Sneden, C., Laird, J. B., Aguilar, L. A., \& Latham, D. W. 1997, AJ, 114, 363
\bibitem[Cayrel et al.(2004)]{Cayrel_04}
Cayrel, R. et al. 2004, A\&A, 416, 1117
\bibitem[Christlieb et al.(2002)]{Christlieb_02}
Christlieb, N. et al. 2002, Nature, 419, 904
\bibitem[Frebel et al.(2005)]{Frebel_05}
Frebel, N. et al. 2005, Nature, 434, 871 
\bibitem[Fujimoto et al.(1984)]{Fujimoto_84}
Fujimoto, M. Y., Iben, I. Jr., Chieffi, A., \& Tornamb\'{e}, A. 1984, ApJ, 287, 749
\bibitem[Fujimoto et al.(2000)]{Fujimoto_00}
Fujimoto, M. Y., Ikeda, Y., \& Iben, I. Jr. 2000, ApJ, 529, L25
\bibitem[Fulbright(2000)]{Fulbright_00}
Fulbright, J. P. 2000, AJ, 120, 1841 
\bibitem[Fulbright et al.(2004)]{Fulbright_04}
Fulbright, J. P., Rich, R. M., \& Castro, S. 2004, ApJ, 612, 447
\bibitem[Gilmore \& Wyse(1998)]{Gilmore_98}
Gilmore, G., \& Wyse, R. F. G. 1998, AJ, 116, 748
\bibitem[Gratton et al.(2003)]{Gratton_03}
Gratton, R. G., Carretta, E., Claudi, R., Lucatello, S., \& Barbieri, M. 2003, A\&A, 404, 187
\bibitem[Iben \& Renzini(1983)]{Iben_83}
Iben, I. Jr., \& Renzini, A. 1983, ARA\&A, 21, 271
\bibitem[Ivans et al.(2003)]{Ivans_03}
Ivans, I. I. et al. 2003, ApJ, 592, 906
\bibitem[Iwamoto et al.(1999)]{Iwamoto_99}
Iwamoto, K., Brachwitz, F., Nomoto, K., Kishimoto, N., Umeda, H., Hix, W. R., \& Thielemann, F.-K. 
1999, ApJS, 125, 439
\bibitem[Iwamoto et al.(2005)]{Iwamoto_05}
Iwamoto, N., Umeda, H., Tominaga, N., Nomoto, K., \& Maeda, K. 2005, Science, 309, 451
\bibitem[King(1997)]{King_97}
King, J. R. 1997, AJ, 113, 2302
\bibitem[McWilliam(1998)]{McWilliam_98}
McWilliam, A. 1998, AJ, 115, 1640
\bibitem[Nakamura \& Umemura(2001)]{Nakamura_01}
Nakamura, F., \& Umemura, M. 2001, ApJ, 548, 19
\bibitem[Nakasato \& Shigeyama(2000)]{Nakasato_00}
Nakasato, N., \& Shigeyama, T. 2000, ApJ, 541, L59
\bibitem[Nissen \& Schuster (1997)]{Nissen_97}
Nissen, P. E., \& Schuster, W. J. 1997, A\&A, 326, 751
\bibitem[Omukai \& Yoshii(2003)]{Omukai_03}
Omukai, K., \& Yoshii, Y. 2003, ApJ. 599, 746
\bibitem[Ryan et al.(1991)]{Ryan_91}
Ryan, S. G., Norris, J. E., \& Bessell, M. S. 1991, AJ, 102, 303
\bibitem[Schneider et al.(2003)]{Schneider_03}
Schneider, R., Ferrara, A., Salvaterra, R., Omukai, K., \& Bromm, V. 2003, Nature, 422, 869
\bibitem[Schuster \& Nissen(1989)]{Schuster_89} Schuster, W.~J., \& 
Nissen, P.~E.\ 1989, \aap, 222, 69 
\bibitem[Shigeyama \& Tsujimoto(1998)]{Shigeyama_98}
Shigeyama, T., \& Tsujimoto, T. 1998, ApJ, 507, L135
\bibitem[Shigeyama et al.(2003)]{Shigeyama_03}
Shigeyama, T., Tsujimoto, T., \& Yoshii, Y. 2003, ApJ. 586, 57
\bibitem[Siess et al.(2002)]{Siess_02}
Siess, L., Livio, M., \& Lattanzio, J. 2002, ApJ, 570, 329
\bibitem[Stephens \& Boesgaard(2002)]{Stephens_02}
Stephens, A., \& Boesgaard, A. M. 2002, AJ, 123, 1647
\bibitem[Suda et al.(2004)]{Suda_04}
Suda, T., Aikawa, M., Machida, M. N., Fujimoto, M. Y., \& Iben, I. Jr. 2004, ApJ, 611, 476 
\bibitem[Tsujimoto et al.(1999)]{Tsujimoto_99}
Tsujimoto, T., Shigeyama, T., \& Yoshii, Y. 1999, ApJ, 519, L63
\bibitem[Tsujimoto et al.(2000)]{Tsujimoto_00}
Tsujimoto, T., Shigeyama, T., \& Yoshii, Y. 2000, ApJ, 531, L33
\bibitem[Umeda \& Nomoto(2003)]{Umeda_03}
Umeda, H., \& Nomoto, K. 2003, Nature, 422, 871
\end{thebibliography}
\end{document}